\documentclass[10pt]{article} 
\usepackage{fullpage}

\usepackage{epsfig}
\usepackage{graphicx}
\usepackage{bm}
\usepackage{amsmath}
\usepackage{amsfonts}
\usepackage{amssymb}
\usepackage{latexsym}
\usepackage{psfrag}
\def\<{\langle}
\def\>{\rangle}

\newcommand{\be}{\begin{equation}}
\newcommand{\ee}{\end{equation}}

\begin{document}

\title{\bf Multi-photon entanglement from distant single photon sources on demand}

\author{Almut Beige,$^1$ Yuan Liang Lim,$^2$ and Christian Sch\"on$^3$ \\[0.3cm]
{\small $^1$The School of Physics and Astronomy, University of Leeds, Leeds, LS2 9JT, United Kingdom} \\[-0.1cm]
{\small $^2$DSO National Laboratories, 20 Science Park Drive, Singapore 118230, Singapore} \\[-0.1cm]
{\small $^3$Blackett Laboratory, Imperial College London, Prince Consort Road, London SW7 2BZ, United Kingdom}}

\date{\today}

\maketitle

\begin{abstract}
We describe a scheme that allows for the generation of
any desired $N$-photon state {\em on demand}. Under ideal conditions, this requires only $N$
single photon sources, laser pulses and linear optics
elements. First, the sources should be initialised with the
help of single-qubit rotations and repeat-until-success two-qubit
quantum gates [Lim {\em et al.}, Phys. Rev. Lett. {\bf 95}, 030305
(2005)]. Afterwards, the state of the sources can be mapped onto
the state of $N$ newly generated photons whenever needed. \\[0.25cm]
\end{abstract}

\section{Introduction}

Photons are natural carriers of quantum information and their states
are favoured for a variety of applications. The reason is that
photons are very robust against decoherence and possess an ease in
distribution. Multi-photon states are therefore the crucial
ingredient for quantum cryptography \cite{Ekert}, quantum
lithography \cite{litho} and for experimental tests of quantum mechanics, like the ones performed to
rule out local hidden variable theories \cite{Clauser,Aspect}.
Photons are also often used in quantum computing experiments \cite{KLM,Browne}. Most of these applications exploit the quantum correlations that are found in entangled states. In
this paper we present a scheme, which can be used to
generate {\em any} desired multi-photon state {\em on demand} and might lead to a variety of previously unconsidered applications of photon entanglement.

Unfortunately, it is very difficult to create an effective
interaction between photons and hence they are difficult to
entangle. Photon entanglement is therefore traditionally generated
by the means of a single source. The first source of this type was the atomic
cascade, which was used for the first test of Bell's inequality \cite{Aspect}. Another example is spontaneous parametric downconversion. It enables the creation of maximally
entangled photon pairs and is routinely used in the laboratory
\cite{Kwiat,Tittel}. However, the generated entanglement is
necessarily probabilistic and can only be detected postselectively.
Hence, parametric down conversion does not scale easily and linear optics
experiments with more than five photons are extremely challenging
\cite{Zhao,Walther}.

New perspectives for multi-photon experiments arise from recent
developments of relatively reliable single photon sources consisting
for example of a single atom \cite{Kuhn2,kimble,grang} or a quantum
dot \cite{cavdot,cavdot2} placed inside an optical
resonator. Using such sources, multi-photon entanglement can be
generated with the help of linear optics elements and postselection
\cite{Lee,Fiurasek02,firework}. However, this approach is not
deterministic and results in the destruction of the photons upon
detection. Avoiding the latter in postselective schemes would
require the use of ready-on-demand entangled multi-photon states as ancillas
\cite{Pittman,hummingbird,Browne}. One possibility to create
multi-photon entanglement in a deterministic manner is to use a
single-photon source with a relatively complex level structure and
to map the state of the source and operations performed on it
subsequently onto the state of newly generated photons
\cite{Gheri,Lange2}. Recently it has been shown by Sch\"on {\em et
al.} \cite{Schon} that this approach can be used to generate
arbitrary multi-photon states on demand.

However, instead of trying to entangle the photons directly, one can
also entangle several single photon sources. Once this is done,
their state can be mapped on demand onto the states of several newly
generated photons whenever needed \cite{LimSPIE,Kok}. In this paper we describe such
a scheme and show that it can be used to generate {\em any} desired
$N$-photon state on demand. Under ideal conditions, this requires only $N$ single
photon sources, which can be initialised with the help of single-qubit rotations and
repeat-until-success two-qubit quantum gates \cite{moonlight}.
Differently from Ref.~\cite{LimSPIE}, we focus on the generation of time-bin instead of
polarisation-entanglement since this decreases the experimental resource requirements significantly.

The presented scheme has many advantages compared to previous proposals for the generation of arbitrary multi-photon entanglement on demand. As in Refs.~\cite{LimSPIE,Kok}, we require only single photon sources,
passive linear optics elements and detectors {\em without} photon number-resolution. Nevertheless, we do not rely on the presence of entangled ancilla photons. The required level structure of the sources is simpler than the ones considered in Refs.~\cite{Schon,Gheri,Lange2,LimSPIE} and has already been tested experimentally (see e.g.~Ref.~\cite{Kuhn2}). Moreover, there is no need for a final step disentangling the states of the sources from the states of the photons with the help of measurements and single-qubit rotations as in Ref.~\cite{Kok}. There are also no restrictions on the location of the photon sources, which can be separated from each other by a long distance, since we do not require an explicit interaction between them.

However, the initialisation step requires high photon creation and detection efficiencies since the repeat-until-success two-qubit gate \cite{moonlight} is otherwise not deterministic. In the presence of photon detectors with a finite efficiency the proposed scheme should be used to prepare the sources in a so-called cluster state \cite{minks}, which can be transformed into any desired state via single-qubit measurements applied to the state of the sources \cite{Raussendorf,Kok}.  Alternatively, but less efficiently, the sources could be prepared in a cluster state with the help of the two-qubit quantum gate by Protsenko {\em et al.} \cite{grangier} or the entangling scheme by Barrett and Kok \cite{barrett}. Once the sources have been initialised, it should be possible to create single photons on demand. This process has realistically only a finite success rate but a read-out measurement performed on the state of the respective source can reveal whether a photon has been emitted or not with a very high efficiency \cite{behe}. 

To generate $N$ entangled photons, the scheme requires at least $N$ single-photon sources. However, these sources do not have to be totally identical hence relaxing what would otherwise be a steep experimental requirement. The repeat-until-success quantum gate is constructed such that each emitted photon contributes equally to each photon detection and path fluctuations between the photon sources and the detectors result at most in an overall phase factor with no physical consequences. The experimental setup is therefore {\em interferometrically stable} and demands only indistinguishability of the sources. The interfering photons do not even need to arrive simultaneously in the detectors as long as they overlap in the setup within their coherence time \cite{Thesis}. To illustrate this point, we start the paper with a discussion of  the photon interference experiment by Eichmann {\em et al.} \cite{Eichmann}.

\section{Photon interference from independent single photon sources}

\begin{figure}
\begin{minipage}{\columnwidth}
\begin{center}
\resizebox{\columnwidth}{!}{\rotatebox{0}{\includegraphics{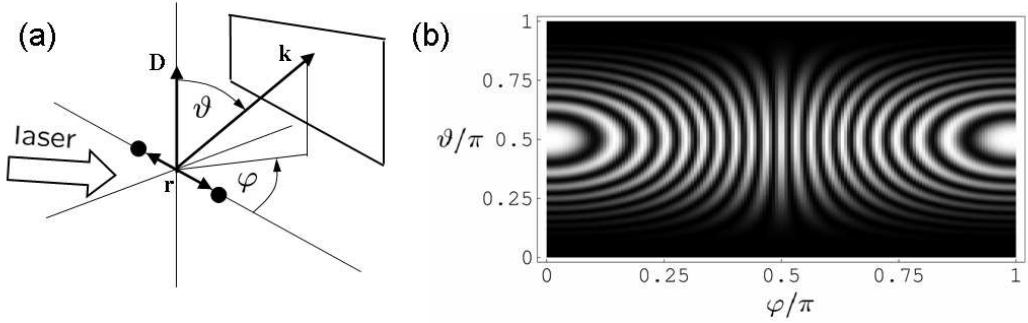}}}
\end{center}
\vspace*{-0.5cm} \caption{(a) Experimental setup. Two two-level
atoms are placed at a fixed distance $r$ from each other. Both are
coupled to the same free radiation field and are continuously driven
by a resonant laser. This leads to spontaneous photon emissions.
Each photon causes a click at a point on a screen. The direction
$\hat{\bf k}$ of the emitted photons is characterized by spatial
angles $\vartheta$ and $\varphi$. (b) Density plot of the emission
rate $I_{\hat{\bf k}}(\rho)$ for two continuously driven two-level
atoms. White areas correspond to spatial angles with maximal
intensity. For the calculation of this figure, we assumed that the
atomic dipole moment ${\bf D}$ is perpendicular to the line
connecting both atoms \cite{Schoen}.} \label{doubleslit}
\end{minipage}
\end{figure}

In 1982, Scully and Dr\"uhl \cite{Scully82} proposed a simple
quantum eraser experiment concerning delayed choice phenomena in
quantum mechanics. The setup they considered is shown in Figure
\ref{doubleslit}(a). It consists of two two-level atoms trapped at a
fixed distance $r$ from each other . The particles are continuously driven by a resonant laser
field and spontaneously emit photons. Each emitted photon causes a
click at a certain point on a screen far away from the particles.
These clicks,  when collected, add up to an interference pattern
with a spatial intensity distribution, found also in
classical double-slit experiments. This was verified experimentally
by Eichmann {\em et al.} in 1993 \cite{Eichmann}. Since then the
interpretation of this experiment attracted a lot of interest in the
literature \cite{Englert,Itano,Schoen}. In this section, we give a
short description of the above described two-atom double-slit
experiment following the discussion by Sch\"on and Beige
\cite{Schoen}.

The time evolution of the quantum mechanical components, namely the
two-level atoms, the surrounding free radiation field and the applied laser light can be
modelled with the help of the Schr\"odinger equation. The role of
the applied laser field is to continuously excite the particles.
Moreover, the interaction between the atoms and the free radiation
field results in the transfer of energy from the excited states of the atoms into the photon modes with wavevector ${\bf k}$ and polarisation $\lambda$. In other words, the Hamiltonian of the
system entangles the state of the atoms with the free radiation
field, whenever there is some population in the excited state.

However, the setup shown in Figure \ref{doubleslit}(a) cannot be
described by the continuous solution of a Schr\"odinger equation. To
take the possibility of spontaneous photon emissions into account,
we also have to consider the environment. The experimental
observation of radiating atoms suggests to model the environment by
assuming rapidly repated measurements whether a photon has been
emitted or not \cite{hegerfeldt}. In case of a click, the direction
$\hat{\bf k}$ of the emitted photon is registered on the screen.
Ref.~\cite{Schoen} showed that the state of two atoms prepared in
$|\psi \rangle$ is, up to normalisation, of the form
\begin{equation} \label{r}
|\psi_{\hat{\bf k}} \rangle \equiv R_{\hat{\bf k}} \, |\psi \rangle  ~~ {\rm with} ~~
R_{\hat{\bf k}} = R_{\hat{\bf k}}^{(1)} + R_{\hat{\bf k}}^{(2)} \, .
\end{equation}
Here $R_{\hat{\bf k}}^{(i)}$ is the reset operator for the case, in
which only atom $i$ is present in the setup, and depends on the
position of the respective atom. Eq.~(\ref{r}) ascertains that
$R_{\hat{\bf k}}$ is the sum of the reset operators of two
independent atoms. This fact is the origin for the presence of an
interference pattern in the two-atom double-slit experiment
\cite{Eichmann}.

We now ask the question,
what is the probability density to observe a click in a certain
direction $\hat{\bf k}$. Under the assumption that the atoms are repeatedly prepared in $|\psi \rangle$, this probability density $I_{\hat{\bf
k}}(\psi) $ is given by the norm squared of the state in
Eq.~(\ref{r}),
\begin{equation} \label{i}
I_{\hat{\bf k}}(\psi) = \| \, R_{\hat{\bf k}} \, |\psi\rangle \,
\|^2 \, .
\end{equation}
Hence this probability density is of the form
\begin{eqnarray} \label{int}
I_{\hat{\bf k}}(\psi) &=& \| \, R_{\hat{\bf k}}^{(1)} \, |\psi
\rangle
+ R_{\hat{\bf k}}^{(2)}  \, |\psi\rangle \, \|^2  \nonumber \\
&=& \| \, R_{\hat{\bf k}}^{(1)} \, |\psi\rangle \, \|^2 + \| \,
R_{\hat{\bf k}}^{(2)} \, |\psi\rangle \, \|^2 + \langle \psi | \,
R_{\hat{\bf k}}^{(1)\dagger} R_{\hat{\bf k}}^{(2)} + R_{\hat{\bf
k}}^{(2)\dagger} R_{\hat{\bf k}}^{(1)} \, |\psi \rangle \, .
\end{eqnarray}
This equation shows that the intensity of the light emitted from two
atoms is not the same as the sum of the light intensities from two
independent atoms (first two terms). The difference is the
interference term (third term), which causes the emission of a photon in
some directions to be more likely than the emission into others. If
one replaces the pure state $|\psi \rangle$ by the stationary state
$\rho$ of the atoms in the presence of continuous laser excitation,
Eq.~(\ref{int}) can be used to calculate the interference pattern
for the experimental setup in Figure \ref{doubleslit}(a). The result is
shown in Figure \ref{doubleslit}(b) and agrees well with the observation
in the experiment by Eichmann {\em et al.} \cite{Eichmann}.

The coupling between the atoms and the free radiation field results
in the transfer of energy into the free radiation field. In general,
there is less than one excitation in each mode $({\bf k}, \lambda)$
of the free radiation field. The interference pattern stems from the
wave behaviour of these excitations prior to the detection of a
photon. Note that a {\em single photon} does not exist until it is
actually detected on the screen. Quantum mechanics tells us that the
observations on the screen always result in the detection of an
integer number of photons. Moreover, each detected photon is created
by both atoms and leaves a trace in and contains information about
all its respective sources.

\section{An entangling two-qubit gate operation between distant single photon sources}

\begin{figure}
\begin{minipage}{\columnwidth}
\begin{center}
\resizebox{\columnwidth}{!}{\rotatebox{0}{\includegraphics{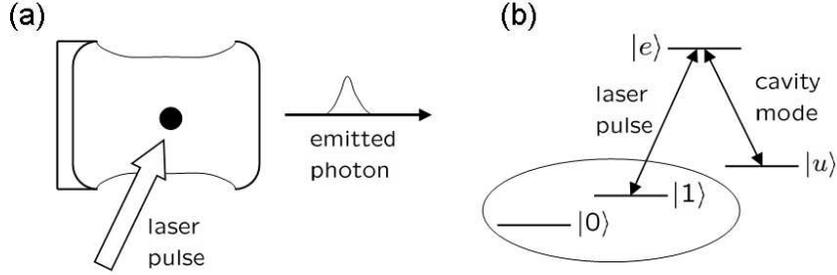}}}
\end{center}
\vspace*{-0.5cm} \caption{(a) Experimental setup. An external laser
pulse drives the atom. The cavity mode decays through the right
lossy mirror and a single-photon pulse builds up in the
corresponding time-bin. (b) Atomic level structure. The two ground
states $|0\>$ and $|1\>$ define the atomic qubit. Level $|1\>$
($|u\>$) and $|e\>$ are coupled by a laser (cavity mode), while
level $|0\>$ is not affected.} \label{pistol}
\end{minipage}
\end{figure}

In the following, we use the possibility of strong correlations
between emitted photons and the state of their respective sources to
process quantum information efficiently. More concretely, we
consider single photon sources consisting of an
atom-like system with a $\Lambda$-level configuration trapped inside
a resonant optical cavity (c.f.~Figure \ref{pistol}). To generate a
single photon on demand, a laser pulse with a relatively slowly
increasing Rabi frequency should be applied. Such a pulse transfers
an atom initially prepared in $|1 \rangle$ via an adiabatic passage
into $|u \rangle$, thereby placing one photon into the cavity
\cite{Kuhn2,kimble}. From there it leaks out through the
outcoupling mirror. Repumping the atom afterwards into its initial
state results in the overall operation
\begin{equation} \label{one}
|1 \rangle ~~ \longrightarrow ~~  |1 ; 1_{\rm ph} \rangle \, .
\end{equation}
The role of the cavity is to fix the direction of the spontaneously
emitted photon so that it can be easily processed further.

Suppose each atom within a large network of single photon sources
contains one qubit consisting of the two ground states $|0 \rangle$
and $|1 \rangle$ (c.f.~Figure \ref{pistol}). Then it is possible to
generate a single photon on demand such that its state depends on
the initial state of the atom and
\begin{equation} \label{two}
\alpha \, |0 \rangle + \beta \, |1 \rangle ~~ \longrightarrow ~~
\alpha \, |0;{\sf E} \rangle + \beta \, |1;{\sf L} \rangle \, .
\end{equation}
Here $|{\sf E} \rangle$ denotes an {\em early} and $|{\sf L}
\rangle$ denotes a {\em late} photon. One way to implement the
encoding step (\ref{two}) is to first swap the atomic states $|0
\rangle$ and $|1 \rangle$. Next a laser pulse with increasing Rabi
frequency should be applied to perform the operation (\ref{one}).
Afterwards, the states $|0 \rangle$ and $|1 \rangle$ should be
swapped again and the photon generation process (\ref{one}) should be
repeated at a later time. In the final state (\ref{two}), the qubit
is double encoded in the state of the source as well as in the state
of the photon. The encoding (\ref{two}) is the main building block for the
realisation of a deterministic entangling gate operation between two
qubits. It requires the simultaneous generation of a photon in each
of the involved single photon sources. Afterwards, the photons
should pass, within their coherence time, through a linear optics
network which performs a photon pair measurement on them. 

Suppose, the two qubits involved in the gate operation are initially prepared
in the arbitrary two-qubit state
\begin{equation} \label{in}
|\psi_{\rm in} \rangle = \alpha \, |00 \rangle + \beta \, |01
\rangle + \gamma \, |10 \rangle + \delta \, |11 \rangle \, .
\end{equation}
Using Eq.~(\ref{two}), we see that the state of the system equals in this case
\begin{equation} \label{enc}
|\psi_{\rm enc} \rangle = \alpha \, |00;{\sf EE} \rangle + \beta \,
|01;{\sf EL} \rangle + \gamma \, |10;{\sf LE} \rangle + \delta \,
|11;{\sf LL} \rangle
\end{equation}
after the creation of the photons. If a photon pair measurement
afterwards results in the detection of a state of the form
\begin{equation} \label{max}
|\Phi \rangle = |{\sf EE} \rangle + {\rm e}^{{\rm i} \varphi_1} \,
|{\sf EL} \rangle + {\rm e}^{{\rm i} \varphi_2} \, |{\sf LE} \rangle
+ {\rm e}^{{\rm i} \varphi_3} \, |{\sf LL} \rangle \, ,
\end{equation}
then the state of the photon sources is projected onto
\begin{equation}
|\psi_{\rm fin} \rangle  = \alpha \, |00 \rangle + {\rm e}^{-{\rm i}
\varphi_1} \, \beta \, |01 \rangle + {\rm e}^{-{\rm i} \varphi_2} \,
\gamma \, |10 \rangle + {\rm e}^{-{\rm i} \varphi_3} \, \delta \,
|11 \rangle \, .
\end{equation}
This state differs from the one in Eq.~(\ref{in}) by a unitary
operation, namely a two-qubit phase gate. As we see below, if the
state (\ref{max}) is a maximally entangled state, this operation is
a phase gate with maximum entangling power.

Performing a deterministic gate operation requires a complete set of
basis states with each of them being of the form (\ref{max}). Such a
basis is called mutually unbiased \cite{Wootters}, since its 
observation does not reveal any information about the coefficients $\alpha$, $\beta$,
$\gamma$ and $\delta$. The limitation of linear optics is that it is
not possible to perform a complete Bell measurement
\cite{Luetkenhaus2}. At most two maximally entangled
photon states can be distinguished using only available linear
optics elements. We therefore consider in the following a measurement of the basis
states \cite{moonlight}
\begin{equation}
|\Phi_{1,2} \rangle = {\textstyle {1 \over \sqrt{2}}} \, \big(
|{\sf x}_1 {\sf y}_2 \rangle \pm |{\sf y}_1 {\sf x}_2 \rangle \big)
\, , ~~ |\Phi_3 \rangle = |{\sf x}_1 {\sf x}_2 \rangle \, , ~~
|\Phi_4 \rangle = |{\sf y}_1 {\sf y}_2 \rangle
\end{equation}
with
\begin{eqnarray}  \label{xy}
&& |{\sf x}_{1,2} \rangle \equiv  {\textstyle {1 \over \sqrt{2}}} \, \big( |{\sf E} \rangle + |{\sf L} \rangle \big) \, , ~~
|{\sf y}_1 \rangle \equiv  {\textstyle {1 \over \sqrt{2}}} \, \big( |{\sf E} \rangle - |{\sf L} \rangle \big)\, , ~~   
|{\sf y}_2 \rangle  \equiv  {\rm i}  \, |{\sf y}_1 \rangle \, .
\end{eqnarray}
This definition implies
\begin{eqnarray} \label{Phi}
&& |\Phi_{1,2} \rangle = \pm {\textstyle {1 \over 2}} \, {\rm e}^{\pm{\rm i} \pi/4}  \, \big( |{\sf EE} \rangle \mp {\rm i} \, |{\sf EL} \rangle \pm {\rm i} \,  |{\sf LE} \rangle - |{\sf LL} \rangle \big) \, , ~~ \nonumber \\
&& |\Phi_3 \rangle =  {\textstyle {1 \over 2}} \, \big( |{\sf EE} \rangle + |{\sf EL} \rangle + |{\sf LE} \rangle + |{\sf LL} \rangle \big) \, , ~~
|\Phi_4 \rangle =  {\textstyle {1 \over 2}} \, {\rm i} \, \big( |{\sf EE} \rangle -  |{\sf EL} \rangle - |{\sf LE} \rangle + |{\sf LL} \rangle \big) \, .
\end{eqnarray}
A comparison with Eq.~(\ref{max}) shows that these $|\Phi_i \rangle$
are indeed mutually unbiased. To find out which quantum gate
operation belongs to which measurement outcome $|\Phi_i \rangle$, we
write the encoded state (\ref{enc}) as
\begin{equation}
|\psi_{\rm enc} \rangle = {\textstyle {1 \over 2}} \, \sum_{i=1}^4 |\psi_i
\rangle |\Phi_i \rangle
\end{equation}
and find that
\begin{eqnarray}
&& |\psi_{1,2} \rangle = \pm {\rm e}^{\mp {\rm i} \pi/4} \, Z_1\big(\pm {\textstyle {1 \over 2}} \pi \big) \, Z_2 \big( \mp {\textstyle {1 \over 2}} \pi \big) \, U_{CZ} \, |\psi_{\rm in} \rangle \, , ~~ |\psi_3 \rangle = |\psi_{\rm in} \rangle \, , ~~ |\psi_4 \rangle = -  {\rm i} \,  Z_1(\pi) \, Z_2(\pi) \, |\psi_{\rm in} \rangle 
\end{eqnarray}
with 
\begin{equation} \label{xxx}
Z_i(\varphi) \equiv {\rm diag} \, (1, {\rm e}^{-{\rm i}\varphi}) \, ,  ~~  U_{\rm CZ} \equiv {\rm diag} \, ( 1, 1, 1, -1 ) \, .
\end{equation}
Here $Z_i(\varphi)$ describes a one-qubit phase gate that changes
the phase of an atom if it is prepared in $|1 \rangle$ and $U_{\rm CZ}$ denotes a controlled two-qubit phase gate with maximum entangling power. The above equations show that a measurement of $|\Phi_1 \rangle$ or
$|\Phi_2 \rangle$ results with probability $1/2$ in the
completion of the universal phase gate $U_{\rm CZ}$ up to local
operations, which can be undone easily. A measurement of $|\Phi_3
\rangle$ or $|\Phi_4 \rangle$ yields the initial qubits up to local
operations. Since the quantum information stored in the system is
not lost at any stage of the computation, the above described steps
can be {\em repeated until success}. On average, the completion of
one repeat-until-success quantum gate requires two repetitions.

\begin{figure}
\begin{minipage}{\columnwidth}
\begin{center}
\resizebox{\columnwidth}{!}{\rotatebox{0}{\includegraphics{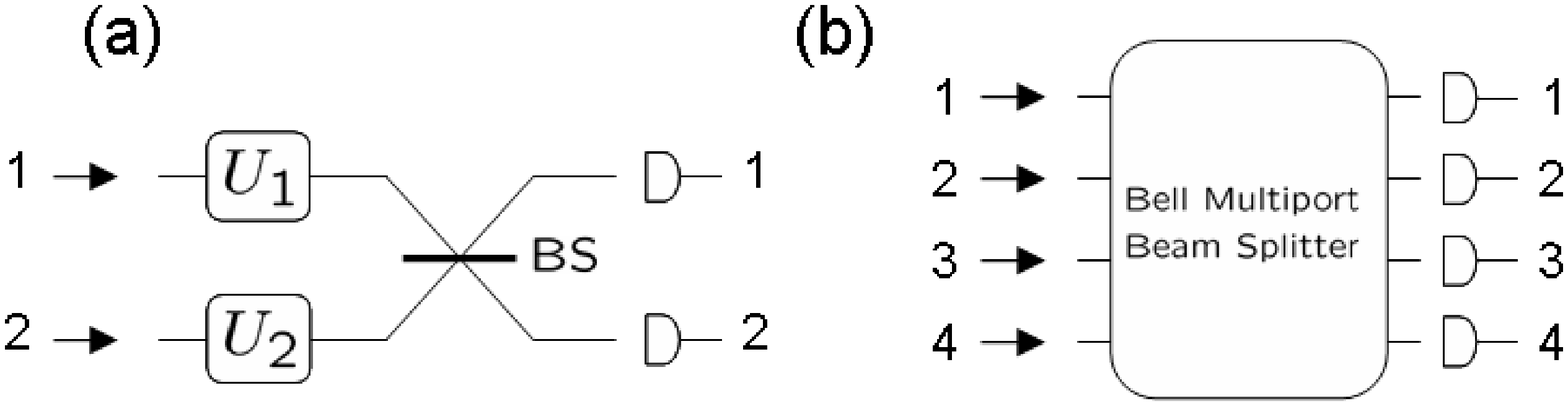}}}
\end{center}
\vspace*{-0.5cm} \caption{Two possible experimental setups for the
realisation of the photon pair measurement in a mutually unbiased
basis.} \label{repeat}
\end{minipage}
\end{figure}

Let us now comment on the experimental feasibility of the proposed
two-qubit gate. One way to realise the above described photon pair
measurement is to convert the time bin encoding of the photonic
qubits into a polarisation encoding. It is known that sending two
polarisation encoded photons through a beam splitter results in a
measurement of the states $|{\sf hv \pm vh} \rangle$, $|{\sf hh}
\rangle$ and $|{\sf vv}\rangle$. Measuring the states $|\Phi_i
\rangle$ therefore only requires passing the photons through a beam
splitter after applying the mapping $U_i = |{\sf h} \rangle \langle {\sf x_i}| + |{\sf v} \rangle \langle {\sf
y}_i|$ on the photon coming from source $i$ with $|x_i \rangle$ and $|y_i
\rangle$ as in Eq.~(\ref{xy}). This is illustrated in Figure
\ref{repeat}(a). Figure \ref{repeat}(b) shows an alternative
approach in which the time bin encoded photons are send through a
Bell multiport beam splitter. An early (late) photon from source 1
should enter input port 1 (3) and an early (late) photon from source
2 should enter input port 2 (4). More details can be found in
Refs.~\cite{moonlight,minks}.

\section{The initialisation of the sources}

For the initialisation of the photon sources we combine the
two-qubit phase gate from above with single-qubit rotations,
achieved by standard quantum optics techniques as routinely employed
in ion trap experiments \cite{blatt,blatt2}. They constitute a universal set of gates and can
therefore be used to prepare $N$ single photon sources in any
desired entangled state~\cite{Nielsen2000}. Especially easy to prepare are $N$-qubit states that belong to the class of two-dimensional matrix-product states. They can be generated using only $N-1$ two-qubit gate operations between ``neighbouring'' sources, since they can always be written as 
\begin{eqnarray} \label{MPS}
|\psi\> = V_{N-1} \, \ldots \, V_1 \, |i_1, \ldots, i_N \> \, .
\end{eqnarray}
Here $|i_n \rangle$ is a one-qubit state and $V_n$ is a two-qubit quantum gate acting on qubit $n$ and qubit $n+1$. Which operations $V_n$ have to be applied in order to prepare a
certain final state is described in detail in Ref.~\cite{Schoen06}. Familiar examples of two-dimensional
matrix-product states are $W$ states, $GHZ$ states and, most importantly, one-dimensional cluster states \cite{VerstCirQC}. To generate a one-dimensional cluster state, the $V_n$ are the controlled-phase gates $U_{\rm CZ}$ defined in Eq.~(\ref{xxx}) if the single photon sources are initially all prepared in $|i_n \rangle = \big( |0 \rangle + |1 \rangle \big)/\sqrt{2}$ \cite{Raussendorf}.

\section{The generation of multi-photon entanglement on demand}

Once the atomic qubits have been initialised, $N$ photons in exactly
the same state can be created by simply mapping the state of the
sources onto the state of $N$ newly generated photons. This can be done whenever
required, since the atoms act as perfect single photon storage devices. The steps involved in this mapping process are similar to
the steps involved in the generation of an encoded photon
(\ref{two}) and can be realised using exactly the same atomic level
configuration (c.f. Figure \ref{pistol}(b)). Suppose the state of a single photon source equals 
\begin{equation}
|\psi_{\rm at} \rangle = \alpha \, |0 \rangle + \beta \, |1 \rangle \, .
\end{equation}
To generate a photon in the same state, a single-qubit rotation should be applied that exchanges the states $|0 \rangle$ and $|1 \rangle$. Afterwards a laser pulse transfers the state $|1 \rangle$ into the state $|u \rangle$,
thereby resulting in the generation of an {\em early} photon. Finally, another exchange of the states $|0 \rangle$ and $|1 \rangle$ and the creation of a {\em late} photon conditional on the atom being in state $|1 \rangle$ results in the overall operation
\begin{equation}
|\psi_{\rm at} \rangle \rightarrow \alpha \, |u;{\sf E} \rangle + \beta \, |u;{\sf L} \rangle \, .
\end{equation}
The final state is a state with the atom in $|u \rangle$ and a
single photon in 
\begin{equation}
|\psi_{\rm ph} \rangle = \alpha \, |{\sf E} \rangle + \beta \,
|{\sf L} \rangle \, .
\end{equation}
The state of the atom has indeed been mapped onto the state of a newly generated photon.

The mapping of the state of $N$ independent single photon sources
onto $N$ newly generated photon can be done in exactly the same way.
Instead of generating a photon from only one source, photons should
be created in each of the prior initialised sources. Given the atoms
are initially prepared in the state
\begin{equation} \label{ooo}
|\psi_{\rm at} \rangle = c_{0 \, \ldots \, 0} \, |0 \, \ldots \, 0 \rangle + c_{0 \, \ldots \, 01}
\, |0 \, \ldots \, 01 \rangle + \, \ldots \, + c_{1 \, \ldots \, 1} \, |1 \, \ldots 1 \, \rangle \, ,
\end{equation}
this results in the generation of the multi-photon state
\begin{equation}
|\psi_{\rm ph} \rangle = c_{0 \, \ldots \, 0} \, |{\sf E} \, \ldots \, {\sf E}
\rangle + c_{0 \, \ldots \, 01} \, |{\sf E} \, \ldots \, {\sf EL} \rangle + \, \ldots \, + c_{1 \, \ldots \, 1} \,
|{\sf L} \, \ldots \, {\sf L} \rangle \, ,
\end{equation}
while the atoms all end up in $|u \rangle$. Here we did nothing else,
than replacing the states $|0 \rangle$ and $|1 \rangle$ in Eq.~(\ref{ooo}) by the
single photon states $|{\sf E} \rangle$ and $|{\sf L} \rangle$. As
mentioned before, the state of the newly generated photons is
exactly the same as the initial state of the $N$ single photon
sources.

\section{Conclusions}

We described a scheme that allows for the generation of any desired entangled $N$-photon state on demand. 
Under ideal conditions, the scheme requires only $N$ single photon sources and passive linear optics elements. If necessary, linear optics can also be employed to transform the time-bin encoded photons into polarisation encoded ones. The basic idea is to first prepare the single photon sources in the desired state. Afterwards their state can be mapped onto the state of $N$ newly generated photons whenever needed.

The initialisation of the sources can be achieved using single-qubit rotations and the repeat-until-success quantum computing scheme by Lim {\em et al.} \cite{moonlight}, which results in the realisation of an eventually deterministic entangling two-qubit phase gate upon the application of a carefully chosen photon pair measurement. In order to elucidate the underlying physical mechanism, we recalled a recent discussion about  interference in the fluorescence of two laser-driven non-interacting atoms~\cite{Schoen}. We also pointed out, that our scheme is especially efficient when engineering two-dimensional matrix-product states, such
as {\em W}-states, GHZ states and one-dimensional cluster states \cite{Schoen06}. 

Considering recent advances on single photon sources and detectors, the two most important ingredients in our scheme, raises hope that an experimental realisation will soon be feasible. When using photon detectors with finite efficiencies and when the photon generation is not ideal, the repeat-until success quantum gate becomes  probabilistic. However, the proposed scheme can still be used to generate multi-photon entanglement on demand. The reason is that even probabilistic quantum gates can be used to prepare the sources efficiently in a so-called cluster state \cite{barrett,minks}, which can then be transformed into the desired state using single-qubit measurements on the states of the sources \cite{Raussendorf}. Whether a photon has been generated in a certain source or not can be detected afterwards by measureing whether the source is prepared in the state $|u \rangle$ or not (c.f.~Figure \ref{pistol}(b)). \\

\noindent {\em Acknowledgment.} AB acknowledges funding from the
Royal Society and the GCHQ as the James Ellis University Research
Fellow. This work was supported in part by the European Union
through the SCALA project and by the UK Engineering and Physical
Sciences Research Council through its Interdisciplinary Research
Collaboration on Quantum Information Processing.


\vspace*{1cm}
\begin{thebibliography}{99}
\bibitem{Ekert}
A. K. Ekert, Phys. Rev. Lett. {\bf 67}, 881 (1991).

\bibitem{litho}
A. N. Boto, P. Kok, D. S. Abrams, S. L. Braunstein, C. P. Williams,
and J. P. Dowling, Phys. Rev. Lett. {\bf 85}, 2733 (2000).

\bibitem{Clauser}
J. F. Clauser, M. A. Horne, A. Shimony and R. A. Holt, Phys. Rev.
Lett, {\bf 23}, 880 (1969).

\bibitem{Aspect}
A. Aspect, P. Grangier, and G. Roger, Phys. Rev. Lett. {\bf 49}, 91 (1982).

\bibitem{KLM}
E. Knill, R. Laflamme, and G. Milburn, Nature {\bf 409}, 46 (2001).

\bibitem{Browne}
D. E. Browne and T. Rudolph, Phys. Rev. Lett. {\bf 95}, 010501 (2005).

\bibitem{Kwiat}
P. G. Kwiat, K. Mattle, H. Weinfurter, A. Zeilinger, A. V. Sergienko, and Y. H. Shih, Phys. Rev. Lett. {\bf 75}, 4337 (1995).

\bibitem{Tittel}
W. Tittel, J. Brendel, H. Zbinden, and N. Gisin, Phys. Rev. Lett. {\bf 81}, 3563 (1998).

\bibitem{Zhao}
Z. Zhao, Y.-A. Chen, A.-N. Zhang, T. Yang, H. J. Briegel, and J.-W. Pan, Nature, {\bf 430}, 54 (2004).

\bibitem{Walther}
P. Walther, K. Resch, T. Rudolph, E. Schenck, H. Weinfurter, V. Vedral, M. Aspelmeyer, and A. Zeilinger, Nature {\bf434}, 169 (2005).

\bibitem{Kuhn2}
A. Kuhn, M. Hennrich, and G. Rempe, Phys. Rev. Lett. {\bf 89}, 067901 (2002).

\bibitem{kimble}
J. McKeever, A. Boca, A. D. Boozer, R. Miller, J. R. Buck, A. Kuzmich, and H. J. Kimble, Science {\bf 303}, 1992 (2004).

\bibitem{grang}
B. Darquie, M. P. A. Jones, J. Dingjan, J. Beugnon, S. Bergamini, Y. Sortais, G. Messin, A. Browaeys, and P. Grangier, Science {\bf 309}, 454 (2005).

\bibitem{cavdot}
J. P. Reithmaier, G. Sek, A. Loffler, C. Hofmann, S. Kuhn, S. Reitzenstein, L. V. Keldysh, V. D. Kulakovskii, T. L. Reinecke, and A. Forchel, Nature {\bf 432}, 197 (2004).

\bibitem{cavdot2}
A. Badolato, K. Hennessy, M. Atature, J. Dreiser, E. Hu, P. M.
Petroff, and A. Imamoglu, Science {\bf  308}, 1158 (2005).

\bibitem{Lee}
H. Lee, P. Kok, N. J. Cerf, and J. P. Dowling, Phys. Rev. A {\bf 65}, 030101 (2002).

\bibitem{Fiurasek02}
J. Fiurasek, Phys. Rev. A {\bf 65}, 053818 (2002).

\bibitem{firework}
Y. L. Lim and A. Beige, Phys. Rev. A {\bf 71}, 062311 (2005).

\bibitem{Pittman}
T. B. Pittman, B. C. Jacobs, and J. D. Franson, Phys. Rev. A {\bf 64}, 062311 (2001).

\bibitem{hummingbird}
Y. L. Lim and A. Beige, J. Mod. Opt. {\bf 52}, 1073 (2005).

\bibitem{Gheri}
K. M. Gheri, C. Saavedra, P. T{\"o}rm{\"a}, J. I. Cirac, and P. Zoller, Phys. Rev. A, {\bf 58}, R2627 (1998).

\bibitem{Lange2}
W. Lange and H. J. Kimble, Phys. Rev. A {\bf 61}, 063817 (2000).

\bibitem{Schon}
C. Sch{\"o}n, E. Solano, F. Verstraete, J. I. Cirac, and M. M. Wolf, Phys. Rev. Lett. {\bf 95}, 110503 (2005).

\bibitem{LimSPIE}
Y. L. Lim and A. Beige, Proc. SPIE {\bf 5436}, 118 (2004).

\bibitem{Kok}
P. Kok, S. D. Barrett and T. P. Spiller, J. Opt. B, {\bf 7}, S166 (2005).

\bibitem{moonlight}
Y. L. Lim, A. Beige and L. C. Kwek, Phys. Rev. Lett. {\bf 95}, 030535 (2005).

\bibitem{minks}
Y. L. Lim, S. D. Barrett, A. Beige, P. Kok, and L. C. Kwek, Phys. Rev. A {\bf 73}, 012304 (2006).

\bibitem{Raussendorf}
R. Raussendorf and H. J. Briegel, Phys. Rev. Lett. {\bf 86}, 5188 (2001).

\bibitem{grangier}
I. E. Protsenko, G. Reymond, N. Schlosser, and P. Grangier, Phys. Rev. A {\bf 66}, 062306 (2002).

\bibitem{barrett}
S. D. Barrett and P. Kok, Phys. Rev. A {\bf 71}, 060310(R) (2005).

\bibitem{behe}
A. Beige and G. C. Hegerfeldt, J. Mod. Opt.  {\bf 44}, 345 (1997). 

\bibitem{Thesis}
Y. L. Lim, {\sl Quantum Information Processing with Single Photons}, PhD Thesis at Imperial College London (2005); quant-ph/0509168.

\bibitem{Eichmann}
U. Eichmann, J. C. Berquist, J. J. Bollinger, J. M. Gilligan, W. M.
Itano, and D. J. Wineland, Phys. Rev. Lett. {\bf 70}, 2359 (1993).

\bibitem{Scully82}
M. O. Scully and K. Dr\"uhl, Phys. Rev. A {\bf 25}, 2208 (1982).

\bibitem{Englert}
B.-G. Englert, Phys. Rev. Lett. {\bf 77}, 2154 (1996).

\bibitem{Itano}
W. M. Itano, J. C. Berquist, J. J. Bollinger, D. J. Wineland, U.
Eichmann, and M. G. Raizen, Phys. Rev. A {\bf 57}, 4176 (1998).

\bibitem{Schoen}
C. Sch\"on and A. Beige, Phys. Rev. A {\bf 64}, 023806 (2001).

\bibitem{hegerfeldt}
G. C. Hegerfeldt and D.G. Sondermann, Quantum Semiclass. Opt. {\bf
8}, 121 (1996).

\bibitem{Wootters}
W. K. Wootters and B. D. Fields, Annals of Physics {\bf 191}, 363 (1989).

\bibitem{Luetkenhaus2}
N. L\"utkenhaus, J. Calsamiglia, and K. A. Suominen, Phys. Rev. A
{\bf 59}, 3295 (1999).

\bibitem{blatt}
F. Schmidt-Kaler, H. H\"affner, M. Riebe, S. Gulde, G. P. T. Lancaster, T. Deuschle, C. Becher, C. F. Roos, J. Eschner, and R. Blatt, Nature {\bf 422}, 408 (2003).

\bibitem{blatt2}
D. Leibfried, B. DeMarco, V. Meyer, D. Lucas, M. Barrett, J. Britton, W. M. Itano, B. Jelenkovic, C. Langer, T. Rosenband, and D. J. Wineland, Nature {\bf 422}, 412 (2003).

\bibitem{Nielsen2000}
M. Nielsen and I. Chuang, {\sl Quantum computation
and quantum information}, Cambridge Univ. Press (2000).

\bibitem{Schoen06}
C. Sch\"on {\ et al.} (in preparation).

\bibitem{VerstCirQC}
F. Verstraete and J. I. Cirac, Phys. Rev. A {\bf 70}, 060302 (2004).
\end{thebibliography}
\end{document}